\documentclass[preprint,aps,prb,groupedaddress,amssymb,showpacs]{revtex4}
\usepackage{graphicx}
\usepackage{dcolumn}
\usepackage{amsmath}
\begin{document}

\title{Far-field image of Veselago lens}

\author{C. Y. Li}
\affiliation{University of Utah, Salt Lake City UT, 84112 USA}

\author{J. M. Holt}
\affiliation{University of Utah, Salt Lake City UT, 84112 USA}

\author{A. L. Efros}
\email{efros@physics.utah.edu}
\affiliation{University of Utah, Salt Lake City UT, 84112 USA}

\date{\today}

\begin{abstract}
It is shown that perfect imaging of a point source both in near- and far-field regions contradicts electrodynamics although ``superlensing'' is impossible only in the far-field region. These general statements are illustrated by detailed study of evanescent wave propagation in a photonic crystal that is known to be a left-handed medium. An analytical expression for the intensity distribution near the far-field focus is obtained. This distribution contains some novel features and establishes a new ``diffraction limit'' for flat lenses. It is generalized for multiple sources located at different points. The distribution is in very good agreement with computer simulation.
\end{abstract}
\maketitle

\section{Introduction}
The left-handed medium (LHM), defined by Veselago \cite{ves} as a medium with simultaneously negative and real $\mu$ and $\epsilon$, recently has attracted much attention mostly because of the negative refraction at its interface with a regular medium (RM). This effect allows creation of a unique device called the ``Veselago lens''. This lens is a slab of LHM inside a RM with a condition that both media have the same {\em isotropic} refractive index and the same impedance. Interest in LHM's significantly increased after the work by Pendry \cite{pen} which argued that the Veselago lens is a ``perfect lens'' in the sense that it gives a perfect image of the point source. This statement is based upon the observation that the evanescent waves (EW's) of a form $\exp (ik_y y -\kappa x)$ that usually decay in the near-field region are amplified by the LHM. Pendry claimed that the amplified EW's restore a perfect image in both the near-field and far-field regions. Following Veselago, Pendry considered a {\em hypothetical} LHM (HLHM) with negative and real $\mu$ and $\epsilon$. He did not present a solution in coordinate space. This was done by Ziolkowski and Heyman\cite{ziol} and their solution reveals a fundamental problem with Pendry's idea. The solution diverges exponentially at each point of a 3-D domain near the focus, just where the fields of the EW's increase due to amplification.  This was found out almost simultaneously by three groups.\cite{gar,pok3,hal} Pokrovsky and Efros\cite{pok3} used diffraction theory, known to be exact at small deviations from geometrical optics, and found a finite width of the focus in the far-field region. Haldane\cite{hal} argued that the problem in general does not have a solution in the framework of macroscopic electrodynamics. 

In this paper, we study the far-field imaging of a two dimensional Veselago lens. In the first section, we present two-dimemsional diffraction theory and use it to find an analytical expression for the intensity near the focal point of the Veselago lens in the HLHM model. As in the three-dimensional case, it is exact in the far-field region and does not suffer from any of the problems introduced by EW's. We present also a general proof that the perfect imaging of a point source contradicts to wave optics. All results obtained in this section are independent of the microscopic structure of the LHM that forms the lens.

In the second section, we consider a uniaxial photonic crystal (PC) and present computer simulations to show that, in the frequency range near the $\Gamma$-point of the second Brillouin zone, it behaves as a real LHM, described by Veselago, with respect to propagating mode\cite{ef, pok} but it does not provide universal amplification of EW's. The reason is that, for the mathematical description of the EW in the PC, one should take into account the spatial dispersion ({\bf k}-dependence) of $\mu$ and $\epsilon$. However, the amplification may (or may not) appear due to some other reasons rather than negative  $\mu$ and $\epsilon$. For example, the amplification might be due to  surface waves that are not related to the dielectric properties of the PC. Such an amplification may provide an improvement of the image in the near-field region, but it does not affect the image near the far-field focal point of the Veselago lens.  Using these results, we provide a wave theory, more general than diffraction theory, which gives exactly the same analytical expression for the field intensity near the far-field focus as diffraction theory. 

In the third section, we present the results of a computer simulation for focusing of both a thin (near-field regime) and thick (far-field regime) Veselago lens and show that the far-field image is described by the analytical theory without the contribution of EWs with very good accuracy, while the near-field image may have ``superlensing'' features. We also define a class of field distributions which gives a perfect far-field image without any ``superlensing''. In fact, the Fourier transform of such fields should not include the EW's at a given frequency.

\section{Two-dimensional diffraction theory and image of the Veselago lens}
The two dimensional Green function of the Helmholtz equation in a RM can be written in a form $G_R=(i/4)H_0^{(1)}(\rho k_0)$, where $k_0= \omega n/c$, the Hankel function $H^{(1)}_0=J_0+iN_0$, and $\rho=\sqrt{(x-x_0)^2+(y-y_0)^2}$, where $x_0$ and $y_0$ are the coordinates of the point source. The function $G_R$  oscillates as $\exp{ik_0\rho}$ at large $\rho$.  The Green function in the LHM should have a form $G_L=-(i/4)H_0^{(1)\ast}(\rho k_0)$.  At large $\rho$, it oscillates as $\exp{-ik_0\rho}$, while at small $\rho$, it has the same form as $G_R$.

We consider a two dimensional problem in the plane $x-y$ with a point source at a distance $a$ from the LHM slab (see Fig. 1). It creates a magnetic field in the $z$-direction. We want to find the field near both foci. It is easy to show that the perfect image of this source is impossible. Indeed, if the image has the same form as the source, the magnetic field near the focus would obey the inhomogeneous Helmholtz equation with a delta-function at the focus, while the focal point would not contain any source. Thus, this solution does not obey the homogeneous Helmholtz equation near the focal point. Note that this argument is general and does not specify the nature of the lens.

\begin{figure}
\includegraphics[width=8.6cm]{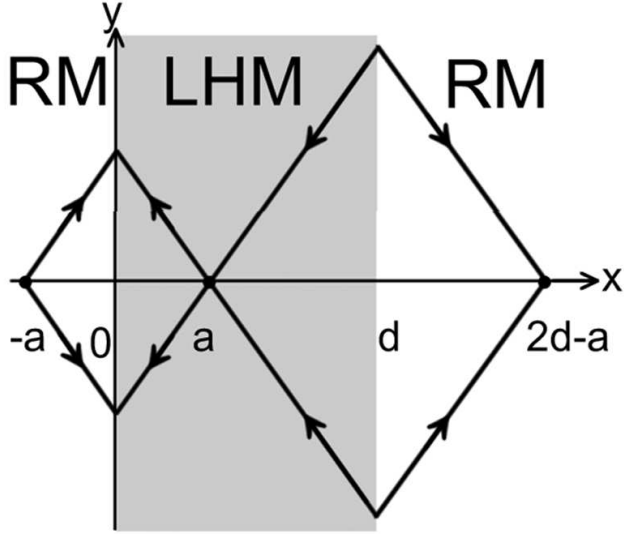}
\caption{Refraction of light outgoing from a point source at $x = -a$ and passing through a slab of LHM at $0 < x < d$ described by anomalous Snell's law. A second focus (the image) is formed at $x = 2d - a$. The arrows show the direction of the wave vector.
\label{fig1}}
\end{figure}

To find the distribution of the field, we use diffraction theory based upon Huygen's principle.\cite{lan} The theory is exact at small wavelength when deviation from geometrical optics is small. Suppose that the field $H(y)$ is given on a line $x=0$. Since the wavelength is small, the field can be represented locally as a plane wave with wave vector $(k_x,k_y)$. The field inside the LHM can be found by implementation of Huygens's principle in the two-dimensional case:
\begin{equation}
H(x,y)=b_{rl}\int_{-\infty}^{\infty}H(y^{\prime})G_L(k_0\sqrt{x^2+(y^{\prime}-y)^2})\cos(\theta)dy^{\prime}.
\end{equation}
Here $\theta$ is the angle between the direction of the ray and the $x$-axis.

To find $b_{rl}$, we use the same method as Ref. 9. Assume that the field to the left of the LHM slab has a form of a plane wave, namely,
 $H=H_0\exp{i k_0x}$. Then Eq. (1) should give us $H=H_0\exp{-i k_0 x}$ inside the LHM slab if $k_0 x\gg 1$. From this condition one can find $b_{rl}$. Since we assume $k_0 x\gg 1$, the Hankel function can be taken in the form 
\begin{equation}
 H_0^{(1)}(k_0\rho) \approx \sqrt{\frac{2}{\pi k_0 \rho}} \exp i(k_0\rho-\frac{\pi}{4}).
\end{equation}
One gets $\cos(\theta)=1$ for a plane wave at normal incidence. To perform the integration in Eq. (1) we can expand the exponent with respect to $y^{\prime}-y$ and neglect $y^{\prime}-y$ in denominator. The result is $b_{rl}=2ki$.

Now we find the field in the LHM created by the point source at $x=-a,$ $y=0$ (see Fig. 1). The field of the source at $x<0$ has a form $H_s(x,y)=iH_0H_0^{(1)}(k_0\rho)$, where $H_0$ is the amplitude and $\rho=\sqrt{(x+a)^2+y^2}$. Thus we should substitute $H(y^{\prime})=H_s(0,y^\prime)$ and $\cos(\theta)=a/\sqrt{a^2+y^{\prime 2}}$ into Eq. (1). Then inside the LHM slab ($0<x<d$) the field has a form
\begin{equation}
 H(x,y) =H_0  \frac{ia}{\pi} \int_{-\infty}^\infty \frac{\exp{ik \left( \sqrt{a^2+y^{\prime 2}}-\sqrt{x^2+(y-y^\prime)^2} \right) }} {(a^2+y^{\prime 2})^{3/4}\left[x^2+(y-y^{\prime})^2\right]^{1/4}} dy^{\prime}.
\end{equation}

If one considers the field near the first focal point ($y=0$, $x=a$) by assuming $ y \ll a $, $|x-a|\ll a$, $ka \gg 1 $ and introducing $\eta_1 = k_0(x-a)$ and $\zeta = k_0y$ one obtains
\begin{equation}
H(x,y) = H_0\frac{i}{\pi} \int_{-1}^1 \frac{\exp{i(\zeta t - \eta_1\sqrt{1-t^2})}}{\sqrt{1-t^2}}dt. 
\end{equation}
To find the field in the RM region $x>d$ we can use the Huygens's principle in the form
\begin{equation}
H(x,y)=b_{lr}\int_{-\infty}^{\infty}H(y^{\prime})G_R(k_0\sqrt{x^2+(y^{\prime}-y)^2})\cos(\theta)dy^{\prime},
\end{equation}
where integration is performed over the right interface.
Using a plane wave, as done above, one finds that the coefficients $b_{rl}=-b_{lr}$. The field $H(y^{\prime})$ at the right interface can be obtained from Eq. (3), by setting $x=d$. In the vicinity of the second focus ($x=2d-a$ and $y=0$) one gets 

\begin{equation}
H(x,y) = H_0\frac{i}{\pi} \int_{-1}^1 \frac{\exp{i(\zeta t + \eta\sqrt{1-t^2})}}{\sqrt{1-t^2}}dt, 
\end{equation}
where $\eta = k_0(x-2d+a)$.

The field in the $y$-direction at $x = 2d-a$ is
\begin{equation}
H(2d-a,\zeta)= H_0\frac{2i}{\pi} \int_0^1 \frac{cos\left(\zeta t\right)}{\sqrt{1-t^2}}dt = iH_0J_0(\zeta)
\end{equation}
and the horizontal profile of the second focus at $y = 0$ is
\begin{equation}
H(\eta,0)= H_0\frac{2i}{\pi} \int_0^1 \frac{\left[cos(\eta q)+isin(\eta q)\right]}{\sqrt{1-q^2}}dq \\
= H_0\left[-{\rm H}_0(\eta) + iJ_0(\eta)\right],
\end{equation}
where $J_0(z)$ and ${\rm H}_0(z)$ are the Bessel and the Struve functions, respectively. The dimensionless intensity $|H(x,y)/H_0|^2$ near the foci is shown in Fig. 2 in both perpendicular ($x$) and lateral ($y$) directions as a function of a distance in units of wavelength $\lambda = 2\pi/k_0$.

\begin{figure}
\includegraphics[width=8.6cm]{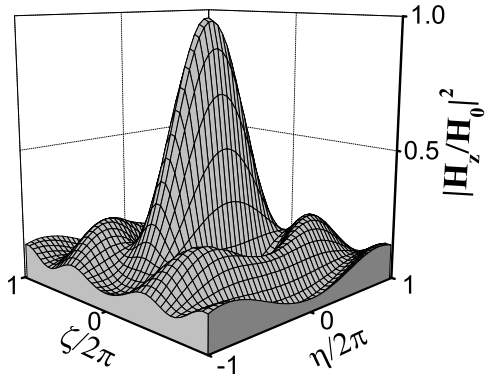}
\caption{Distribution of dimensionless magnetic energy near the foci of the Veselago lens as a function of $\zeta$ and $\eta$ as given by Eq. 6.
\label{fig2}}
\end{figure}

The intensity near the focus is not symmetric with respect to $x$ and $y$: the width of the distribution in the direction perpendicular to the slab is larger than in the lateral direction though the source field depends only on $\rho$.

Another important feature of this imaging is that the image in the lateral direction is  $H(2d-a,y)=iH_0J_0(|y| k_0)\exp{-i\omega t}$. If we consider real fields, the part of the source field that is proportional to $\sin(\omega t)$, namely $H_0J_0(\rho k_0)\sin(\omega t)$, has a perfect image in the lateral direction while the  part of the source with $\cos(\omega t)$ has a zero image in this direction.  This follows only from the flat (non-axial) geometry of the lens that is assumed to be infinite in the lateral direction.  The perfect lateral imaging of the Bessel function $J_0$ does not contradict to general principles because this Bessel function obeys the homogeneous Helmholtz equation.

\section{Propagation of evanescent waves through a left-handed material made of photonic crystal}

We consider a dielectric photonic crystal (PC) that is known to be a LHM in some frequency range.\cite{ef, pok} Here we show that its electrodynamics differs from that of the HLHM, namely that the negative $\mu$ and $\epsilon$ of propagating waves do not provide amplification of EW's.

A large group of works pioneered by Notomi \cite{not} and Luo {\it et al.}\cite{joan,joan2} claims observation of negative refraction, focusing, and even superlensing in a PC due to photons with wave vectors deep in the Brillouin zone. Propagation of such photons can be described macroscopically only by taking into account spatial dispersion (${\bf k}$-dependence of $\epsilon$ and $\mu$). The theory of Veselago can hardly be generalized for the case of spatial dispersion mostly because of an extra term in the definition of Poynting's vector.\cite{lan1} Following Ref. 11, we do not consider such a medium left-handed and do not discuss it here.

It was shown recently,\cite{ef, pok} that a 2-D dielectric uniaxial PC made of non-magnetic materials can behave as a LHM with negative $\epsilon$ and $\mu$ if it has a negative group velocity in the vicinity of the $\Gamma$-point. This was proved for propagating modes only. Experimental demonstration of negative refraction in a metallic PC using the modes near the $\Gamma$-point has been presented in Ref. 14. 

In this paper, we consider p-polarization for both EW's and propagating modes in a uniaxial PC. Fig. 3 shows the elementary cell in the plane $x-y$ and the spectrum of p-polarized propagating waves. The dashed line shows our working frequency ($\omega d/2\pi c=0.33$) for all data below. Computations are performed using the method of finite elements with the software ``FEMLAB''.

\begin{figure}
\includegraphics[width=8.6cm]{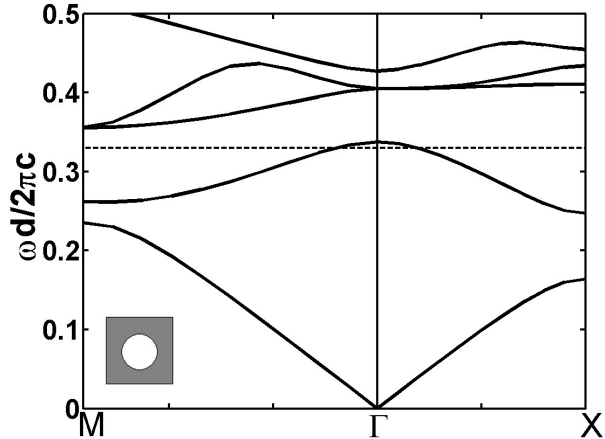}
\caption{Five lowest bands of the photonic spectrum of a 2-D PC with a period $d$. The unit cell of the PC is shown in the inset. The PC consists of a square lattice of circular cylindrical air holes in a dielectric matrix with $\epsilon_m=12$, $\mu_m=1$. The radii of the holes $R=0.35d$. 
\label{fig3}}
\end{figure}

From the same numerical solution of the microscopic Maxwell's equation that gives the spectrum, one gets microscopic fields $h_z(x,y)$, $e_y(x,y)$, and $e_{x}(x,y)$ corresponding to a given frequency of propagating modes. The microscopic fields are the Bloch functions while the macroscopic fields have a form ${\bf B_z}=<$$h_z$$>$, ${\bf E}=$$<$${\bf e}$$>$, where $<$...$>$ means averaging over the unit cell. Near the $\Gamma$-point, the Bloch functions have a small ${\bf k}$ and the macroscopic fields of the propagating modes are plane waves that obey the macroscopic Maxwell equations.

As follows from Ref. \cite{ef}, the values of $ \mu_{zz}\equiv \mu$ and $\epsilon_{xx}=\epsilon_{yy}\equiv \epsilon$ that describe propagation of these waves are negative because group velocity is negative. Here, $z$ is the axis of the crystal. It has been shown previously\cite{ef} that the values of $\epsilon$ and $\mu$ can be expressed through the integrals of Bloch's function at $k=0$. It is important to understand that, in a system with spatial dispersion, magnetization and displacement currents are inseparable.\cite{lan1} Our parameters $\epsilon(\omega)$ and $\mu(\omega)$ are not exclusive properties of the medium at a given frequency, as in macroscopic electrodynamics without spatial dispersion. Rather, they are properties of a particular mode of this medium, whose fields are used for the calculations. We argue below that they are not applicable to the EW's. The theorem\cite{pok} connecting the signs of both $\epsilon(\omega)$ and $\mu(\omega)$ with the sign of group velocity is also not applicable to the EW's. Near the $\Gamma$-point $\mu\sim k^2$.\cite{ef} Since the dispersion law $\omega(k)$ is also isotropic, one obtains $\mu$ that is a function of $\omega$ only. This approach is valid for an infinite PC, but may create problems with boundary conditions \cite{agr} because, in coordinate-space, $k^2$ should be considered as a Laplacian operator. We will refer to this problem later on. For EW's, this approach fails completely because their dispersion law is anisotropic.

To demonstrate the Veselago lens, one should know $\mu$ and $\epsilon$ of the RM medium around the PC. In this paper, we find the value of the refractive index $n=\sqrt{\epsilon\mu}$ from the function $\omega(k)$, shown in Fig. 3, using the definition $\omega=kc/n$. The second condition that provides matching of impedance is the absence of reflection of the incident plane wave in a wide range of incident angles.

Using these two conditions, we get that, at the working frequency, the RM material should have
\begin{equation}
 \epsilon = 1.125, \mu = 0.08.
\end{equation}
The physical reasons for the appearance of $\mu$ in a non-magnetic PC are discussed in Ref. 7. If not stated otherwise, we use these values for the homogeneous regions surrounding the PC slab in the computations below. 

Fig. 4 shows the result of our simulation of light propagation through a PC slab surrounded by a homogeneous medium. One can see the negative refraction of  light  coming in and going out of the PC with equal absolute values of incident and refracted angles and without any visible reflections from the interfaces. We verified this condition over a wide range of the incident angles (from 0 to 60 degrees). An animation \cite{web} shows that the wave fronts are moving to the right outside the PC slab and to the left inside the slab. All of these phenomena exactly correspond to Veselago's theory. \cite{ves}

\begin{figure}
\includegraphics[width=8.6cm]{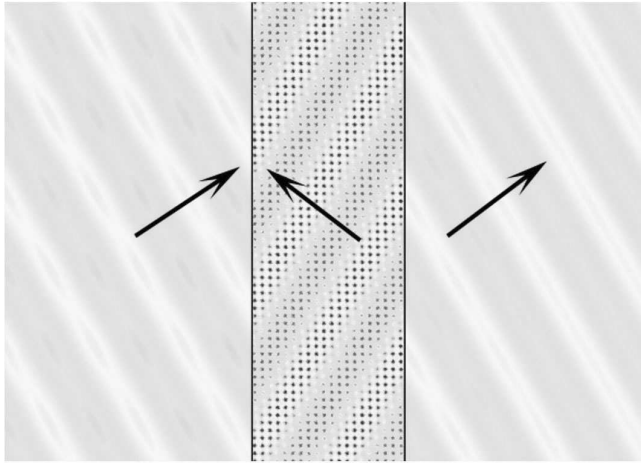}
\caption{Negative refraction of light propagating through a PC slab surrounded by a homogeneous medium at the working frequency. Periodic boundary conditions are used in the vertical direction. The arrows show the direction of the wave vector.
\label{fig4}}
\end{figure}

Now we consider EW's in the same PC slab embedded in the same RM as in Fig. 4. Our results are shown in Fig. 5. The boundary conditions implied for an incident wave at $x=0$ are $h_z = \exp{(i k_y y)}$ and $d h_z / dx = -\kappa h_z$. Note that, due to reflection, the total field shown in Fig. 5 may not coincide at $x=0$ with the incident one. To obey Maxwell's equation, we imply the dispersion law of the EW's in a form $k_y^2-\kappa^2=\omega^2 n^2/c^2$. The distance from the left boundary of the RM (point $x=0$) to the PC slab is 1/2 of the slab thickness. In Fig. 5(a) the PC slab is substituted by the HLHM that has negative $\epsilon$ and $\mu$ with the same absolute values as in the RM. One can see a strong amplification of the EW first discovered by Pendry.$^2$ However, the PC slab gives a very different picture. First, there are oscillations inside the PC with the period $d$ of the PC. The result for the envelope strongly depends on the way the surface is cut. Our surfaces are perpendicular to the [10] direction, but they are cut either across the holes (AH) or between the holes (BH) (see insert in Fig. 5 (b),(c)).

\begin{figure}
\includegraphics[width=8.6cm]{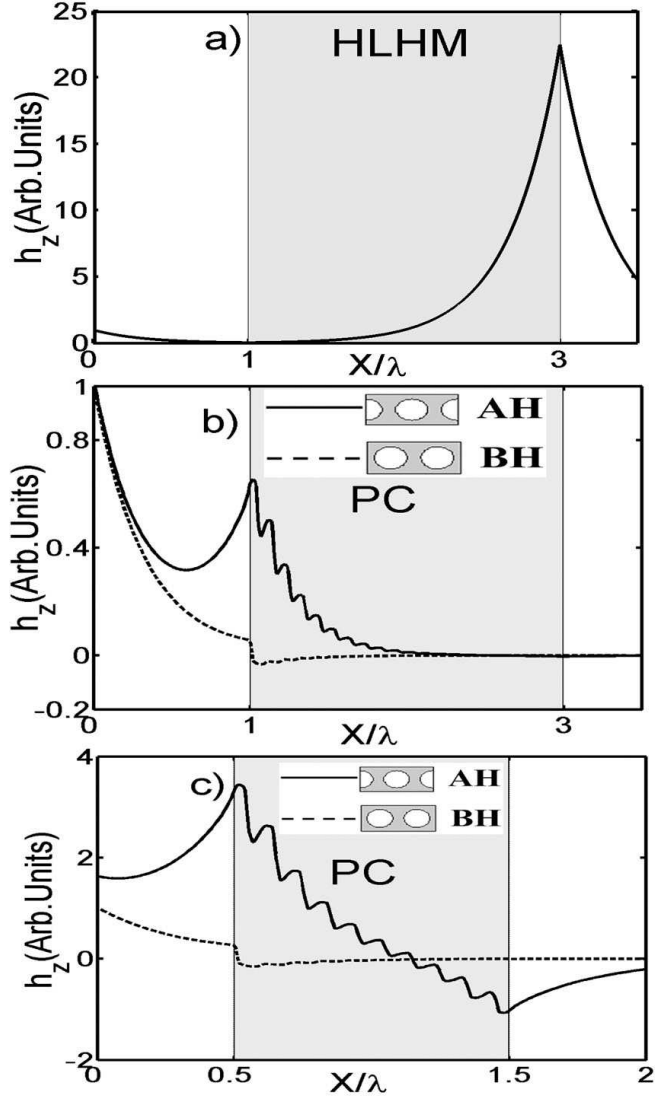}
\caption{Magnetic field of the EW in a HLHM (a) and in a PC slab (b),(c) surrounded by a regular homogeneous medium at the following parameters for the EW: $k_y = (\sqrt{5}/2) k_0$, $\kappa=0.5k_0$, $k_0=\omega n/c$, $n=\sqrt{\epsilon \mu}=0.3$. The cross section $y=0$ that goes across the holes is shown. The results are similar if the cross-section $y=0$ goes between the holes. The thickness of the PC slab is $20d\approx 2\lambda$ (b) and $10d\approx \lambda$ (c). Here $\lambda=2\pi/k_0$ is the wavelength. The solid lines are for the AH surfaces, the dashed lines are for the BH surfaces. The period of oscillation is $d$.
\label{fig5}}
\end{figure}

In the case of the BH surface (dashed lines), the EW decays in the PC slab in the same way as in the RM without any signature of amplification. However, there is amplification at the left AH surface for the thick slab and at both AH surfaces for the thin slab. We believe that these amplifications are due to surface waves (SW's), excited by EW's. Since the picture is not symmetric with respect to left and right interfaces, this excitation is not resonant. These SW's have nothing in common with left-handed properties or with the surface polaritons in the HLHM, considered by Ruppin \cite{rup} because they depend on the properties of the surface, while the polaritons should have a wavelength larger than the lattice constant, so that their spectra depend on the $\epsilon$ and $\mu$ of the bulk materials. We think these SW's appear solely due to the termination of the periodic structure. They have been studied before by many authors \cite{joan2,jo,ha} with and without connection to focusing.

Thus, we have found that EW's cannot be described in terms of the same negative $\epsilon$ and $\mu$ that are found for the propagating waves. The formal reason is that the dispersion law for the EW is anisotropic even near the $\Gamma$-point. Therefore, in this case, $\mu$ cannot be represented in a ${\bf k}$-independent form as in the case of propagating waves. The physical explanation reads that since the EW does not have a Poynting vector in the direction of decay, it is the same for both the regular and the backward wave. Therefore, in this direction, there is no difference between LHM and RM. We have verified, however, (see Fig. 6) that if an EW propagates (has a real $k$) in a direction not parallel to the surface of the PC slab, it exhibits a negative refraction, similar to that shown in Fig. 4.

\begin{figure}
\includegraphics[width=8.6cm]{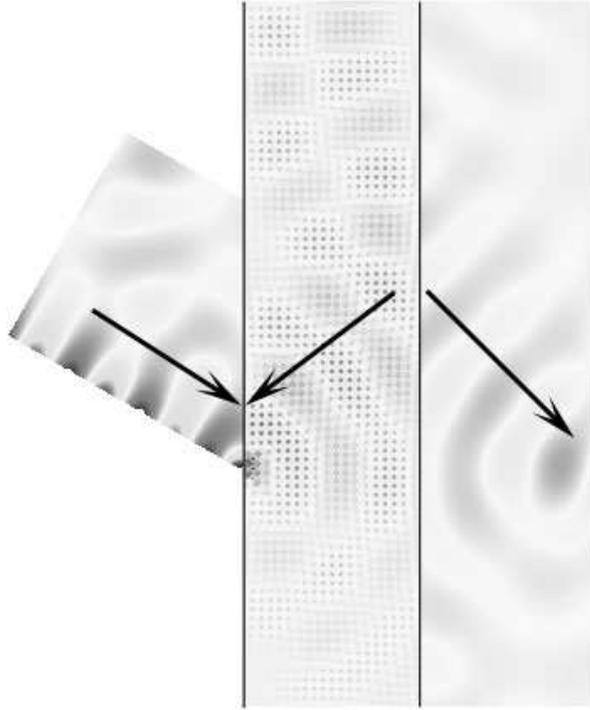}
\caption{An EW propagating in a direction not parallel to the surface of the PC slab exhibits negative refraction at the interface. The arrows show the directions of the propagating wave vector. The EW decays in the plane $x-y$ in the direction perpendicular to the propagating wave vector. 
\label{fig6}}
\end{figure}

Finally, we present a direct proof that the $\epsilon$ and $\mu$ that describe the EW's are ${\bf k}$-dependent and they are {\em positive} at the same frequency where $\epsilon$ and $\mu$ for propagating modes are {\em negative}  and ${\bf k}$-independent. 

First, we make two additional comments about Fig. 5(a):\\
1) The plot Fig. 5(a) is completely the same if {\em the slab of the RM is embedded into the HLHM} and the materials are matched, i.e., they have the same absolute values of $\epsilon$ and $\mu$.\\
2) If the interfaces are not matched, the ``reflected EW'' would appear and there would be a maximum at the left boundary of the HLHM slab.

To simplify the picture, we consider a PC slab with a BH surface so that surface waves are absent. The idea is to find $\epsilon'$ and $\mu'$ of the PC slab that are responsible for the propagation of EW's by choosing $\epsilon$ and $\mu$ of the homogenious regions in such a way that ``reflected EW'' is absent.

Fig. 7 shows propagation of EW's through the PC slab surrounded by a HLHM (Fig. 7(a)) and RM (Fig. 7(b)). To avoid a maximum at the left boundary of the PC slab in Fig. 7(a)  the HLHM should have $\epsilon'=-1.78$ and $\mu'=-0.051$. After that, one can see that the only difference between Figs. 7(a) and 5(a) is oscillations corresponding to the period of the PC. Thus, we get that $\epsilon'=1.78$ and $\mu'=0.051$ are responsible for the propagation of the EW's at a given $k_y$ and $\kappa$ through the PC. Fig. 7(b) shows propagation of a EW through the PC slab surrounded by a RM with $\epsilon'=1.78$ and $\mu'=0.051$. One can see that the EW decays inside the PC as $\exp (-\kappa x)$ with the same value of $\kappa$ as in the RM.

\begin{figure}
\includegraphics[width=8.6cm]{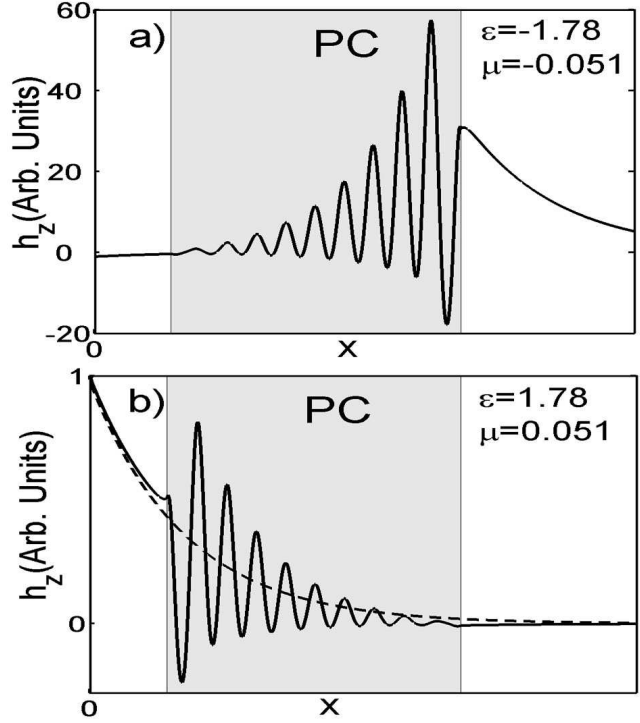}
\caption{Magnetic field of the evanescent wave with $k_y = (\sqrt{5}/2) k_0$, $\kappa=0.5k_0$, $k_0=\omega n/c$ is shown with BH surface for the PC slab. a) The PC slab of the width $10d$ is surrounded by HLHM. b) The same PC slab is surrounded by regular media. The dashed line is a plot of the function $\exp{(-\kappa x)}$. 
\label{fig7}}
\end{figure}

It follows from Fig. 7 that PC slab behaves as a RM with positive values of $\epsilon'$ and $\mu'$ at the same working frequency where propagating modes have negative $\epsilon$ and $\mu$. Moreover, we found that if the values of $\kappa$ and $k_y$ of the EW are changed, keeping the same frequency, the values of $\epsilon'$ and $\mu'$ should be also changed. However, we still retain $\epsilon'\mu' = n^2$. Thus, the parameters of the PC responsible for propagation of EW's are ${\bf k}$-dependent.

\section{Image of the Veselago lens}

Now we consider the image of the Veselago lens. The point source, as provided by our software, is the 2-D Green function $H_s=iH_0H_0^{(1)}(\rho k_0)\exp{-i\omega t}$, where $k_0 = \omega n/c$ and the Hankel function $H^{(1)}_0=J_0+iN_0$. It is located in the RM at a distance $a$ to the left of the PC slab. Similar to the geometry of Fig. 5, the distance $a$ is 1/2 of the thickness of the PC slab. We simulated the Veselago lens with the same PC slab with AH cut. The parameters of the RM around the slab are obtained for the propagating modes and are given by Eq. (9).
  
First, we show that the near-field image obtained with a thin lens is sharper than the far-field image of a thick lens (see Fig. 8). While the far-field image coincides with diffraction theory, the near-field image is beyond the diffraction limit. This is definitely connected with the amplification of EW's by the surface modes, as has been shown in the previous section and, following Pendry, this effect can be called ``superlensing''. Fig. 8 shows the distributions of magnetic energy near the focus of both a thick lens and a thin lens in the lateral ($y$) direction. The distribution is symmetric for $y$ and $-y$ so it is shown for positive $y$ only. Fig. 9 shows the distributions of magnetic energy near the focus of the thick lens in both lateral $(y)$ and perpendicular $(x)$ directions.

\begin{figure}
\includegraphics[width=8.6cm]{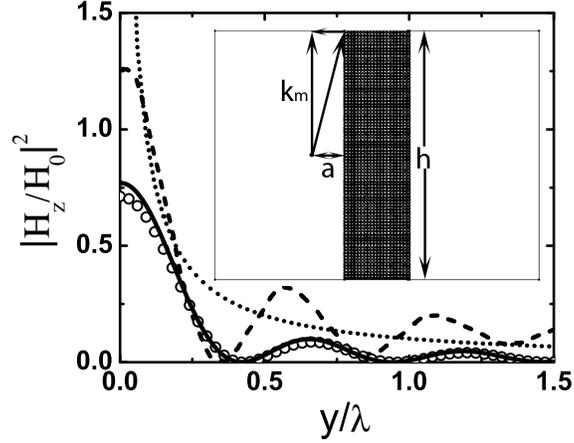}
\caption{Distribution of magnetic energy near the focus both for the thin lens and thick lens with AH surfaces in the lateral direction. The dashed line is the distribution of magnetic energy in the lateral direction for a thin lens; the solid line is the same for a thick lens. Analytical results as obtained from Eq. (13) for a thick lens are shown by circles. The dotted line shows the distribution of magnetic energy of the point source. The inset depicts the geometric meaning of $k_m$. 
\label{fig8}}
\end{figure}

\begin{figure}
\includegraphics[width=8.6cm]{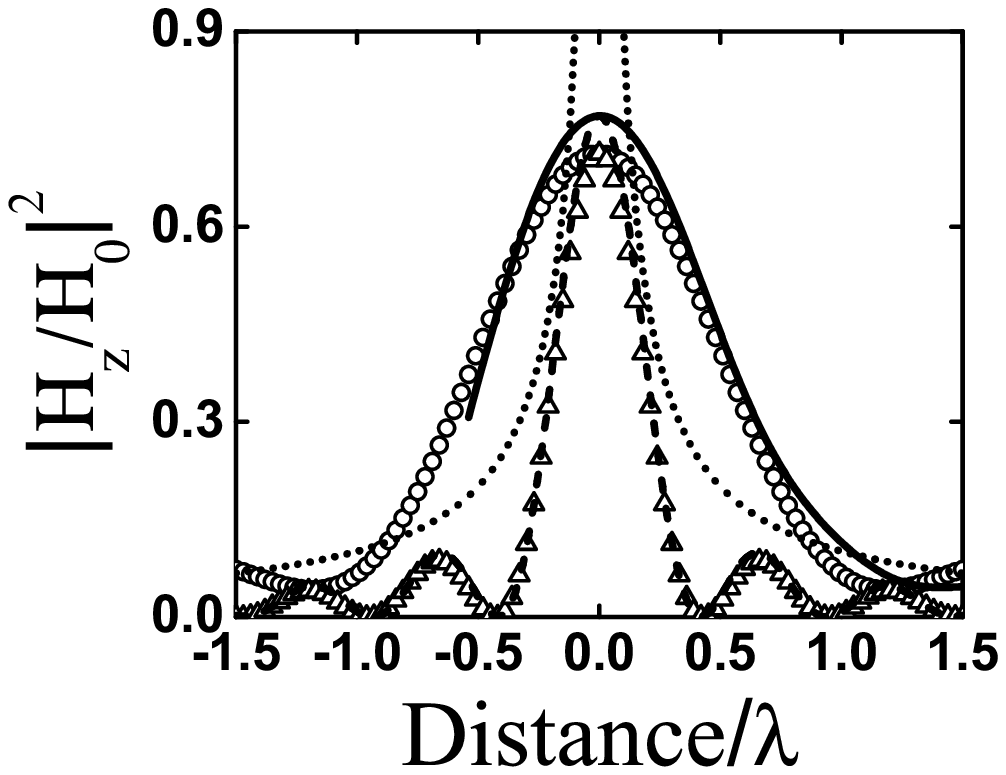}
\caption{Distribution of magnetic energy near the focus for the thick lens with AH surfaces in lateral and perpendicular directions. The solid line is the distribution of magnetic energy in the perpendicular direction; the dashed line is the same for the lateral direction. Circles show the analytical result for perpendicular direction, triangles show the same for lateral direction. Dotted line shows the distribution of magnetic energy of the point source. Analytical results are obtained from Eq. (13).
\label{fig9}}
\end{figure}

Now we discuss the far-field image. The results of Sec. 3 give a possibility to calculate analytically the distribution of energy near the far-field focus of a thick Veselago lens with both AH and BH surfaces or of a thin lens with BH surface. In all these cases, evanescent waves do not reach the focus. At positive $x$, the Green function of the source can be represented in a form $H_s=H_p+H_{ev}$, where 

\begin{equation}
H_p=i(H_0/\pi)\int_{-k_0}^{k_0} \frac{\exp i\left(k y+x\sqrt{k_0^2-k^2}-\omega t\right)}{\sqrt{k_0^2-k^2}}dk 
\end{equation}
contains only propagating modes while 
\begin{equation}
H_{ev}=(H_0/\pi)\int_{|k|>k_0} \frac{\exp\left(ik y-x\sqrt{k^2-k_0^2}-i\omega t\right)}{\sqrt{k^2-k_0^2}}dk 
\end{equation}
contains only EW's.

If the EW's decay inside the PC slab, only term $H_p$ contributes to the focus if the slab is thick enough. On the other hand, our results show that, for the propagating modes, the PC slab works exactly like a HLHM. Using the Fresnel equation, it is easy to show \cite{ziol} that the field near the focus $H_f$ is given by
\begin{equation}
 H_p^f=i(H_0/\pi)\int_{-k_0}^{k_0} \frac{\exp i\left(k y+x^{\prime}\sqrt{k_0^2-k^2}-\omega t\right)}{\sqrt{k_0^2-k^2}}dk.
\end{equation} 
Near the focus, the function $H_f$ exactly coincides with Eq. (6) obtained in Sec. 2 by diffraction theory. Since diffraction theory in the far-field region is exact, it follows that EW's do not contribute to the far-field focus of Veselago lens independent of the microscopic nature of the LHM.

All these results are obtained for the infinite lens. In our simulation, however, the slab has a finite length $h$ in the $y$-direction. In this case, some propagating modes do not enter the slab. To take into account the finite aperture, one should integrate from $-k_m$ to $k_m$ in Eq. (12), where $k_m=k_0(h/2a)/\sqrt{1+(h/2a)^2}$ and $a$ is a distance from a source to a slab (see the inset of Fig. 8). Thus, 
\begin{equation}
H_p^f=i(H_0/\pi)\int_{-k_m}^{k_m} \frac{\exp i\left(k y+x^{\prime}\sqrt{k_0^2-k^2}-\omega t\right)}{\sqrt{k_0^2-k^2}}dk. 
\end{equation}

In our case $h/2a=4,8$ for thick and thin lenses, respectively. Due to this factor, the lateral distribution slightly differs from the square of the Bessel function $J_0^2$ that follows from Eq. (7) for the infinite aperture. The results obtained from Eq. (13) are shown in Fig. 8 for thick (20d) lens. For the thick lens (far-field imaging), analytical calculations are in very good agreement with the computational data. Thus, Eq. (12) sets the diffraction limit for far-field imaging by a  Veselago lens. In Fig. 10, we show the distribution along the lateral direction of two lenses with the same thickness $a=20d$ but a different $h$ of $120d$ and $80d$ correspondingly. As $h$ becomes larger, the normalized intensity distribution approaches the square of the Bessel function. 

\begin{figure}
\includegraphics[width=8.6cm]{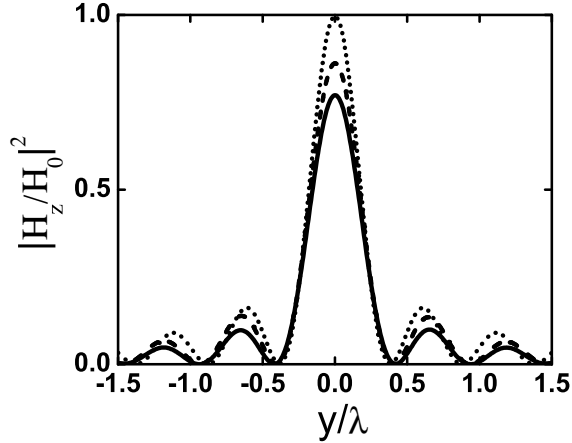}
\caption{Simulation results for the distribution of magnetic energy near the focus in the lateral direction for two lenses with AH surfaces with different heights. The solid line is the distribution for the lens with $h=80d$ and the dashed line is for the lens with $h=120d$. The dotted line is the plot of $J_0^2(2\pi y/\lambda)$. The analytical results are almost indistinguishable from the simulation results. LIfig10.eps
\label{fig10}}
\end{figure}

\begin{figure}[htbp]
\includegraphics[width=8.6cm]{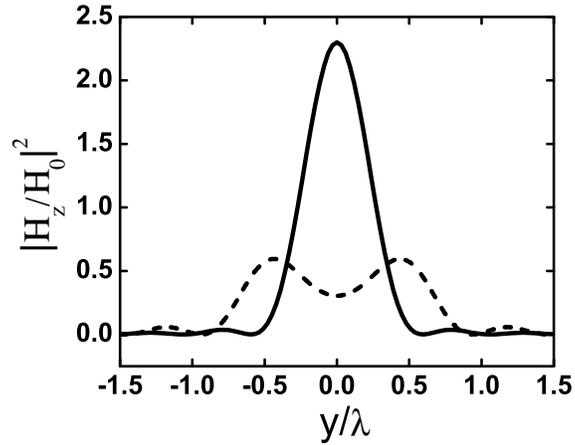}
\caption{Distribution of magnetic energy along the lateral direction near the focus of three point sources with interval $0.3\lambda$ with three point sources located at $y=0$, $y=-0.3\lambda$ and $y=0.3\lambda$, respectively (solid line)  and the same for the interval $0.5\lambda$ with three point sources located at $y=0$, $y=-0.5\lambda$ and $y=0.5\lambda$, respectively (dashed line). The analytical results are almost indistinguishable from the simulation results.
\label{fig11}}
\end{figure}

\begin{figure}
\includegraphics[width=8.6cm]{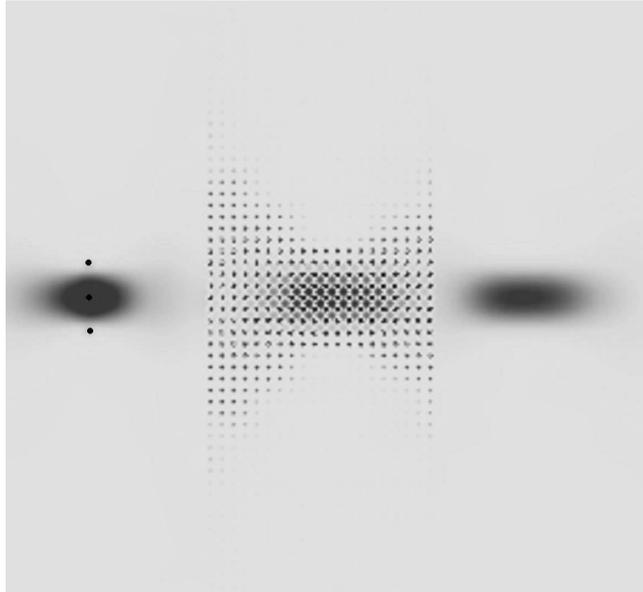}
\caption{Distribution of magnetic energy of three point sources located along the y-axis with interval $0.3\lambda$.  The  distance between the sources and the PC slab is $1/2$ of the slab thickness.
\label{fig12}}
\end{figure}

The first root of the Bessel function is at $y=0.38\lambda$. Thus, the sharp first peak of the  Bessel-like function in the lateral direction can be confused with ``superlensing''. We think, however, that any result that follows from  Eq. (12) has nothing to do with ``superlensing''. Following Pendry, we would define superlensing as a result of amplification of EW's, but Eq. (10) does not contain EW's at all. It gives just the diffraction pattern of the Veselago lens that was first obtained in Ref. 5 for a 3-D lens. Then the energy distribution given by this equation should be called ``regular lensing''.

The superposition principle is valid for combination of point sources with a different location. It can be used to create a focus with {\em one maximum}. To demonstrate this, we put three point sources along $y$ axis with interval $0.3\lambda$ successively, with the middle point source located at $y=0$. In Fig. 11, we show both computational and analytical distribution of energy along the lateral direction near the focus and find that the image contains only one maximum value with the focal point at $y=0$. The same picture shows that three point sources with separation $0.5\lambda$ are already distinguishable by the Veselago lens. Fig. 12 shows the spatial distribution of the field intensity through all the system in the case of three point sources with interval $0.3\lambda$.

An interesting feature of the far-field imaging follows from the above consideration. If the distribution of the object's field does not contain EW's, its far-field image created by an infinite Veselago lens will be perfect. The class of such field distributions includes all functions that can be represented in a form 
\begin{equation}
H(x,y)= \int_{-k_0}^{k_0}F(k)
 \exp{i\left(k y+x\sqrt{k_0^2-k^2}-\omega t\right)}dk,
\end{equation}  
where $F(k)$ is any integrable function in the interval $[-k_0, k_0]$. The function $H(x,y)$ obeys the homogeneous Helmholtz equation. Examples of such functions that produce perfect lateral imaging at $x=0$ are $\sin (k_0y)/y$, $|y|^{-n} J_n(k_0|y|)$ and many others.

\section{Conclusion}
   
In conclusion, we show that a perfect image of the point source is impossible in both near-field and far-field regions and that the Veselago lens does not provide any superlensing in the far-field region. Both statements are independent of the nature of the LHM. We show that the far-field image has some peculiar features, such as perfect imaging for a whole class of functions. We present an analytical expression for the far-field imaging that sets a diffraction limit for the
 Veselago lens. We present an extensive study of the evanescent waves in a photonic crystal showing that a description of EW's can be done in the framework of a theory that includes the spatial dispersion of $\epsilon$ and $\mu$. As a result, both the universal amplification of EW's and the surface polaritons do not exist in the PC. However, in the near-field region (thin lenses) the amplification is possible if the surface supports regular surface waves. For the PC under study a thin lens with AH surface gives a substantially sharper image in the near-field region than that which follows from our diffraction limit. This is due to amplification of EW's shown in Fig. 5(c) and, following Pendry, this effect can be considered as a ``superlensing''. However, the physics of this superlensing is connected with the properties of the surface rather than with the left-handed properties of the medium.
\begin{acknowledgments}
The authors are grateful to E. I. Rashba and S. Luryi for fruitful discussions.
The work has been funded by the NSF grant DMR-0102964 and by the Seed grant of the University of Utah. 
\end{acknowledgments}

\bibliography{image}

\end{document}